\documentstyle[bo99,epsfig]{article}

\title{ASCA view on High-Redshift Radio-Quiet Quasars}

\author{
C. Vignali$^{1,2}$, 
A. Comastri$^{2}$, 
M. Cappi$^{3,4}$, 
G.G.C. Palumbo$^{1}$, 
M. Matsuoka$^{5}$
}

\affil{
1) Dipartimento di Astronomia, Universit\`a di Bologna, 
via Ranzani 1, I--40127, Bologna, Italy \\
2) Osservatorio Astronomico, via Ranzani 1, I--40127, Bologna, Italy \\
3) ITeSRE, C.N.R., via Gobetti 101, I-40129 Bologna, Italy \\
4) Harvard-Smithsonian Center for Astrophysics, 
60 Garden Street, Cambridge MA 02138, USA \\
5) National Space Development Agency of Japan (NASDA), World Trade Center Bldg., 2-4-1, 
Hamamatsu-cho, Minato-ku, Tokyo 105-8060, Japan
}
\begin{document}

\maketitle

\begin{abstract}
We briefly discuss the latest {\it ASCA} results on the 
X--ray spectral properties of high-redshift radio-quiet quasars. 
%%We also suggest some possible scenarios for the observed properties of this class of objects. 
%%
%We also compare these properties with the ones of low-redshift samples and 
%high-redshift radio-loud ones. 
%%
\keywords{active galaxies; quasars; nonthermal mechanism; X-rays}
\end{abstract}

\section{Introduction}
%%
%%The origin of the X--ray emission in quasars and in active galactic nuclei 
%%(AGNs) is likely to be strongly associated with accretion processes 
%%onto a supermassive black hole. It is well known that X--ray 
%%observations allow to investigate the innermost regions of AGNs and to study the 
%%primary emission mechanism responsable for the large energy output. 
%%
The study of quasar X--ray properties as a function of 
redshift can address some important issues such as {\bf (a)} the history of 
accretion processes over the cosmic time, {\bf (b)} the evolution of quasar 
activity and {\bf (c)} the condition 
of the Universe at the epoch of quasars formation. 
So far, while high-redshift radio-loud quasars (RLQs) have been widely studied in hard 
X--rays in the last few years (Cappi et al. 1997, Yamazaki et al. 1998), 
the X--ray properties of high-z radio-quiet quasars (RQQs) are still poorly known 
(even though they constitute about 90 \% of the quasar population), due to their weaker 
X--ray flux with respect to radio-loud objects (Zamorani et al. 1981). 
In order to fill this gap we have performed a pilot study with {\it ASCA} of a 
sample of high-redshift (z $>$ 1.85) RQQs. 

\begin{table}[h]
\footnotesize
\centerline{\bf Table.~1 - The Radio-Quiet Quasars sample}
\begin{center}
\begin{tabular}{|l|c|c|c|c|c|c|}
\hline
\multicolumn{1}{l}{\bf Object} &
\multicolumn{1}{c}{\bf z} &
\multicolumn{1}{c}{$N_{{\rm H}_{\rm gal}}^{a}$} &
\multicolumn{1}{c}{$m_{\rm V}$} &
\multicolumn{1}{c}{Exp.$_{{\rm SIS}{(GIS)}}$} &
\multicolumn{1}{c}{CR$_{{\rm SIS}{(GIS)}}$} &
\multicolumn{1}{c}{$\alpha_{\rm ox}$$^{b}$} \\
\multicolumn{1}{l}{} &
\multicolumn{1}{c}{} &
\multicolumn{1}{c}{} &
\multicolumn{1}{c}{} &
\multicolumn{1}{c}{(ks)} &
\multicolumn{1}{c}{(10$^{-2}$ c/s)} &
\multicolumn{1}{c}{} \\
\hline
{\bf 0151$-$4046} & {\bf 1.85} & 2.07 & 18.1 & 32 (36) & 4.0 (2.3) & 1.15 \\
{\bf 0040$+$0034} & {\bf 2.00} & 2.45 & 18.0 & 30 (36) & 3.8 (3.2) & 1.17 \\
{\bf 1352$-$2242} & {\bf 2.00} & 5.88 & 18.2 & 31.4 & 1.7 (1.5) & 1.29\\
{\bf 1247$+$267} & {\bf 2.04} & 0.9 & 15.6 & 36 (34) & 1.3 (1.1) & 1.69 \\
{\bf 1400$+$10}   & {\bf 2.07} & 1.81 & & 86.5 & 2.4 (1.8) & \\
{\bf 1101$-$264}  & {\bf 2.15} & 5.68 & 16.0 & 17.4 (19.7) & 1 (0.6) & 1.71 \\
{\bf 0059$-$304A} & {\bf 2.17} & 2.00 & 19.3 & 31 (37.4) & 0.5 (0.6) & 1.23 \\
{\bf 0300$-$4342} & {\bf 2.30} & 1.83 & 19.2 & 37.7 & 1.3 (1.2) & 1.16 \\
{\bf 0130$-$4124} & {\bf 2.46} & 2.20 & 20.8 & 35.3 (38) & 1.2 (1.1) & 0.95 \\
\hline
\end{tabular}
\end{center}
\vspace{-0.2cm}
\hspace{1cm} {
$^{a}$ {\small In units of 10$^{20}$ cm$^{-2}$, Dickey \& Lockman 1990}; 
$^{b}$ {\tiny $\alpha_{\rm ox}$ = $-$$\frac{Log(F_{\rm v}/F_{\rm x})}{Log(\nu_{\rm v}/\nu_{\rm x})}$}
}
\end{table}

\section {The sample}
The objects presented here have been chosen among the brightest ones found through a cross 
correlation of the V{\'e}ron-V{\'e}ron quasars catalogue 
(V{\'e}ron-Cetty \& V{\'e}ron 1996) with the ROSAT All-Sky Survey catalogue of X--ray sources 
(Voges et al. 1999). 
We have obtained relatively good ASCA spectra for 9 sources (see Table~1 for the relevant data). 
%%
%one of which retrieved from the {\it ASCA} 
%public archive. 
%%
The sample is clearly not complete and, due to its soft X--ray selection, 
may be biased toward less absorbed objects. 
Nonetheless, it may be considered adequate in order to 
obtain, for the first time and prior to {\it XMM} launch, 
a reliable measurement of the X--ray spectral properties of high-z RQQs. 
A standard analysis has been applied to the data, taking into account 
the most recent calibration uncertainties, and extensive checks on background subtraction have been 
performed. 
%%
%%Moreover, extensive checks on the 
%%influence of background subtraction 
%%on the derived spectral parameters have been applied. 
%We are therefore rather confident that the obtained spectral 
%parameters are a good description of the data. 
%%
A detailed description of the data analysis and the first results 
can be found in Vignali et al. (1999) on a subsample of 4 objects, plus WEE~83, 
which is now excluded from discussion since its redshift has been re-measured and the revised value 
puts it fairly near (z=0.311, Wu et al. 1999). 

\section{Results}
The results of the spectral analysis are presented in Table~2 and summarized in the following. 

\begin{table}[h]
\scriptsize
\centerline{\bf Table~2 - ASCA (SIS$+$GIS) 0.8--10 keV Spectral Results}
\begin{center}
\begin{tabular}{|l|c|c|c|c|c|c|}
\hline
\multicolumn{1}{l}{\bf Object} &
\multicolumn{1}{c}{$N_{\rm H}$} &
\multicolumn{1}{c}{$\Gamma$} &
\multicolumn{1}{c}{R$^{a}$} &
\multicolumn{1}{c}{$\chi^{2}$/dof} &
\multicolumn{1}{c}{$F_{2-10 keV}$} &
\multicolumn{1}{c}{$L_{2-10 keV}$} \\
\multicolumn{1}{l}{} &
\multicolumn{1}{c}{(10$^{21}$ cm$^{-2}$)} &
\multicolumn{1}{c}{} &
\multicolumn{1}{c}{} &
\multicolumn{1}{c}{} &
\multicolumn{1}{c}{(10$^{-13}$ cgs)} &
\multicolumn{1}{c}{(10$^{46}$ cgs)} \\
\hline
{\bf 0040$+$0034} & $\equiv$$N_{{\rm H}_{\rm gal}}$ & 1.66$^{+0.07}_{-0.06}$ 
& & 240/237 & 15 & 2.7 \\
 & $<$ 8.76 & 1.69$^{+0.10}_{-0.08}$ & & 239/236 & & \\
 & $\equiv$$N_{{\rm H}_{\rm gal}}$ & 1.71$^{+0.18}_{-0.14}$ & $<$ 2.34 & 239/236 & & \\
\hline
{\bf 0300$-$4342} & $\equiv$$N_{{\rm H}_{\rm gal}}$ & 1.64$^{+0.12}_{-0.15}$ & & 
168/148 & 5.0 & 1.2 \\
 & $<$ 7.15 & 1.63$\pm{0.14}$ & & 168/147 & & \\
 & $\equiv$$N_{{\rm H}_{\rm gal}}$ & 1.96$^{+0.19}_{-0.33}$ & 3.62$^{+6.38}_{-2.90}$ & 
164/147 & & \\
\hline
{\bf 1101$-$264} & $\equiv$$N_{{\rm H}_{\rm gal}}$ & 1.79$^{+0.26}_{-0.25}$ & & 29.3/23 
& 2.9 & 0.7 \\
& $<$ 63.1 & 2.05$^{+0.62}_{-0.43}$ & & 27.9/22 & & \\
& $\equiv$$N_{{\rm H}_{\rm gal}}$ & 1.70$^{+1.69}_{-0.23}$ & unc. & 29.2/22 & & \\
\hline
{\bf 1352$-$2242} & $\equiv$$N_{{\rm H}_{\rm gal}}$ & 1.66$\pm{0.12}$ & 
& 145/147 & 6.7 & 1.1 \\
 & $<$ 7.53 & 1.65$^{+0.14}_{-0.10}$ & & 145/146 & & \\
 & $\equiv$$N_{{\rm H}_{\rm gal}}$ & 1.84$^{+0.80}_{-0.28}$ & $<$ 23 & 144/146 & & \\
\hline
{\bf 1400$+$10} & $\equiv$$N_{{\rm H}_{\rm gal}}$ & 1.64$^{+0.06}_{-0.05}$ & & 
292/256 & 9.4 & 1.8 \\
 & $<$ 7.64 & 1.64$^{+0.09}_{-0.05}$ & & 292/255 & & \\
 & $\equiv N_{{\rm H}_{\rm gal}}$ & 1.60$^{+0.14}_{-0.08}$ & $<$ 1.13 & 292/255 & & \\
\hline
{\bf 0130$-$4124} & $\equiv N_{{\rm H}_{\rm gal}}$ & 1.78$\pm{0.14}$ & & 
132/137 & 5.0 & 1.6 \\
 & $<$ 54.3 & 1.83$^{+0.32}_{-0.18}$ & & 131/136 & & \\
 & $\equiv N_{{\rm H}_{\rm gal}}$ & 1.69$^{+0.38}_{-0.16}$ & $<$ 2.35 & 132/136 & & \\
\hline
{\bf 0059-304~A} & $\equiv N_{{\rm H}_{\rm gal}}$ & 1.73$^{+0.61}_{-0.55}$ & 
& 88/60 & 2.3 & 0.6 \\
 & $<$ 349 & 1.74$^{+1.79}_{-0.56}$ & & 88/59 & & \\
 & $\equiv N_{{\rm H}_{\rm gal}}$ & 1.92$^{+0.88}_{-0.77}$ & unc. & 87.8/59 & & \\
\hline
{\bf 0151$-$4046} & $\equiv N_{{\rm H}_{\rm gal}}$ & 1.83$^{+0.07}_{-0.06}$ & & 
163/197 & 11 & 2.0 \\
 & $<$ 3.97 & 1.84$\pm{0.07}$ & & 163/196 & & \\
 & $\equiv N_{{\rm H}_{\rm gal}}$ & 1.88$^{+0.19}_{-0.15}$ & $<$ 2.68 & 163/196 & & \\
\hline
{\bf 1247$+$267} & $\equiv N_{{\rm H}_{\rm gal}}$ & 2.00$^{+0.15}_{-0.13}$ & & 
130/124 & 4.8 & 1.3 \\
 & $<$ 2.99 & 2.01$\pm{0.14}$ & & 130/123 & & \\
 & $\equiv N_{{\rm H}_{\rm gal}}$ & 2.23$^{+0.40}_{-0.30}$ & 2.87$^{+7.42}_{-2.69}$ 
& 128/123 & & \\
\hline
\end{tabular}
\end{center}
Note: errors are quoted at 90 \% confidence level for one interesting parameter.\\
$^{a}$ R represents the normalization of reflected vs. direct continuum (R=1 means $\Omega$=2$\pi$ coverage). 
\end{table}

$\bullet$ The spectra are well fitted by a single power law model over the $\sim$ 
2--30 keV energy range (rest frame). The average spectral slope is $<$$\Gamma$$>$ = 1.75, 
with dispersion $\sigma$ = 0.12, which agrees very well with the value ($\Gamma$ = 1.72$\pm{0.03}$) 
obtained from the co-added GIS$+$SIS spectra of all the sources in 
the overlapping rest-frame energy range ($\sim$ 2.4--28.4 keV). 

%--------------------------  figure 1
\begin{figure}
\centerline{\psfig{file=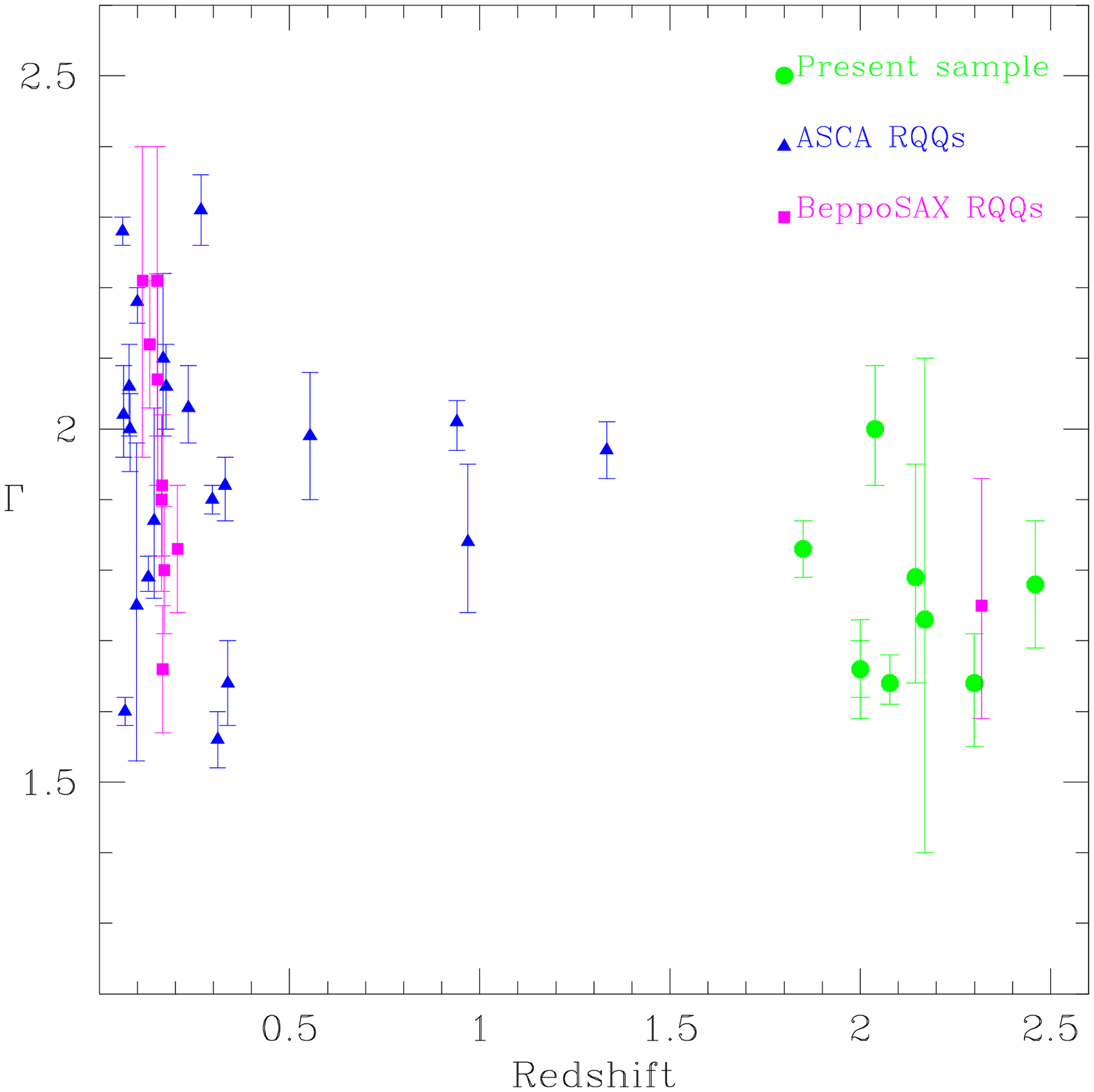, width=7cm}}
\caption[]{RQQs photon spectral indices as a function of redshift.} 
%The {\it ASCA} data are taken from Reeves et al. (1997), George et al. (2000) and the present sample, {\it BeppoSAX} data 
%from Costantini (1998).} 
\end{figure}
%%%
%%%
%$\bullet$ In order to get a higher counting statistics a co-added SIS$+$GIS spectrum of all the 
%sources in the overlapping rest-frame energy range ($\sim$ 2.4--28.4 keV) 
%has been computed. The best-fitting spectral slope is $\Gamma$ = 1.72$\pm{0.03}$. 
%%%
$\bullet$ A comparison with various samples of lower-z RQQs from {\it ASCA} (Reeves et al. 1997, George et al. 2000) 
and {\it BeppoSAX} (Costantini 1998) is showed in Fig.~1, suggesting either a possible flattening of the power law 
slope with redshift or toward high energies. 

$\bullet$ There is no evidence of the signatures of cold matter either in transmission (the upper limits 
on intrinsic absorption ranging from 3 up to 8 $\times$ 10$^{21}$ cm$^{-2}$ rest frame) 
or in reflection, with upper limits on Fe K$\alpha$ line EW 
of $\sim$ 70--200 eV (rest frame). The lack of intrinsic absorption in high-z RQQs 
is at variance with the findings by Elvis et al. (1994) and Cappi et al. (1997) 
for high-z RLQs. If this result will be confirmed by future observations, 
then a different evolution for RQQs and RLQs, possibly related to their environment, 
could be envisaged. 

$\bullet$ A further check on the presence of a reflection component (peaking at 20--30 keV 
rest frame) has been performed on the co-added spectrum. 
The addition of this component results in a steeper spectrum ($\Gamma$ = 1.82$^{+0.11}_{-0.09}$) 
and a value for R (the normalization of the reflected vs. the direct continuum) of 0.89$^{+0.81}_{-0.56}$. 
However, this component statistically is not required by the data. This result, combined with the 
lack of any iron line, do indicate that in high-luminosity RQQs reprocessing, if present, is different 
with respect to nearby Seyfert galaxies. 
%%
%If this component (which, however, remains statistically not significant) was really present,  
%a fluorescence iron K$\alpha$ line should be present as well, but it's not observed in our sample.
%Fixing the photon index to 1.9 (the ``canonical'' value found for nearby AGNs, Nandra \& Pounds 1994) 
%we found R = 1.46$\pm{0.26}$, but this component remains statistically not significant. 
%This means that even in high-luminosity RQQs reprocessing is not working in the same way as in nearby 
%lower-luminosity Seyfert galaxies. 
%%

$\bullet$ The lack of a reflection component and absorption features in the spectra of high-z RQQs 
is not surprising though. It can be explained as the result of 
a strong ionized reflection, which gives rise to 
a reprocessed spectrum similar to the incident one; 
%%
%alternatively, as {\bf (b)} the consequence of 
%an optically-thin accretion disc (in which case the reflection efficiency should be smaller), 
%%
alternatively, it may be caused by a low covering fraction of the reprocessing 
matter as seen from the X--ray source. 
The iron line may be weak due to resonant trapping and Auger effects or totally absent 
if the iron is fully stripped of electrons. 
%%
%The contemporary absence of any iron feature 
%can be explained if iron is fully stripped of electrons or if 
%resonant trapping and Auger effects are destroying line photons. 
%In this case it shoud be difficult to detect this 
%feature by the present detectors like those in {\it ASCA}. 
%%
%{\it Chandra} and {\it XMM} are suitable to test the presence of reprocessing in 
%highly-ionized environments, such as those characterizing the quasars. 
%%

\begin{acknowledgements}
We thank the {\it ASCA} team, who operate the satellite and maintain the software and 
database. Financial support from Italian Space Agency under the contract ASI--ARS--98--119 
and the Italian Ministry for University and Research (MURST) under grant Cofin98-02-32 
are acknowledged by C.~V. and A.~C. 
%%
%This work was supported by the Italian Ministry 
%for University and Research (MURST) under grant Cofin98-02-32. 
%%
\end{acknowledgements}

\end{document}